\documentclass[%
 reprint,
superscriptaddress,
 amsmath,amssymb,
 aps,
 pra,
]{revtex4-2}

\usepackage{lineno}
\usepackage{graphicx}
\usepackage{dcolumn}
\usepackage{bm}

\usepackage{mathrsfs,color,bm,physics}
\newcommand{\eref}[1]{Equation~\ref{#1}}
\newcommand{\fref}[1]{Figure~\ref{#1}}
\newcommand{\sref}[1]{Section~\ref{#1}}
\newcommand{\aref}[1]{Appendix~\ref{#1}}
\newcommand{\innprod}[2]{\left\langle#1,\,#2\right\rangle}
\newcommand{\iu}{{i\mkern1mu}}

\begin{document}

\preprint{APS/123-QED}

\title{Resolvent-Based Optimisation for Approximating the\\Statistics of a Chaotic Lorenz System}

\author{Thomas Burton}
\affiliation{
 Aerodynamics and Flight Mechanics Research Group\\
 University of Southampton
}
\author{Sean Symon}%
\affiliation{
 Aerodynamics and Flight Mechanics Research Group\\
 University of Southampton
}
\author{Ati S Sharma}
\affiliation{
 Aerodynamics and Flight Mechanics Research Group\\
 University of Southampton
}
\affiliation{Agalmic Ltd.}

\author{Davide Lasagna}%
 \email{davide.lasagna@soton.ac.uk}
\affiliation{
 Aerodynamics and Flight Mechanics Research Group\\
 University of Southampton
}

\date{\today}   

\begin{abstract}
We propose a novel framework for approximating the statistical properties of turbulent flows by combining variational methods for the search of unstable periodic orbits with resolvent analysis for dimensionality reduction. Traditional approaches relying on identifying all short, fundamental unstable periodic orbits to compute ergodic averages via cycle expansion are computationally prohibitive for high-dimensional fluid systems. Our framework stems from the observation in \mbox{Lasagna, Phys.~Rev.~E (2020)}, that a single unstable periodic orbit with a period sufficiently long to span a large fraction of the attractor captures the statistical properties of chaotic trajectories. Given the difficulty of identifying unstable periodic orbits for high-dimensional fluid systems, approximate trajectories residing in a low-dimensional subspace are instead constructed using resolvent modes, which inherently capture the temporal periodicity of unstable periodic orbits. The amplitude coefficients of these modes are adjusted iteratively with gradient-based optimisation to minimise the violation of the projected governing equations, producing trajectories that approximate, rather than exactly solve, the system dynamics. A first attempt at utilising this framework on a chaotic system is made here on the Lorenz 1963 equations, where resolvent analysis enables an exact dimensionality reduction from three to two dimensions. Key observables averaged over these trajectories produced by the approach as well as probability distributions and spectra rapidly converge to values obtained from long chaotic simulations, even with a limited number of iterations. This indicates that exact solutions may not be necessary to approximate the system's statistical behaviour, as the trajectories obtained from partial optimisation provide a sufficient ``sketch'' of the attractor in state space.


\end{abstract}

\maketitle


\section{Introduction}\label{sec:introduction}
The high sensitivity to initial conditions coupled with the high number of dynamically significant degrees of freedom pose significant challenges to the detailed prediction of the evolution of turbulent flows. Nonetheless, global behaviour and long-time statistical quantities may be characterised more conveniently by using the notion of a turbulent attractor, a low-dimensional object determining the long-time evolution of turbulent trajectories
\cite{ruelle1971}. The geometry of the attractor may be quite complex and is often fractal in nature \cite{aizawa1982}, allowing solutions to explore it in a complex fashion. For low-dimensional chaotic systems, insight into such geometry may be obtained by examining unstable period orbits (UPOs) densely embedded within the attractor and providing a skeleton that supports the dynamics in state space \cite{eckmann1985}. Using these UPOs in the form of a weighted sum, a technique known as cycle expansion, the ergodic averages of the dynamical system can be computed \cite{cvitanovic1987,cvitanovic1988,cvitanovic1995}.
For turbulent flows, numerical evidence has been found for the existence of time periodic solutions of the Navier-Stokes equations, starting with the first identification of a nonlinear equilibrium flow by Nagata \cite{nagata1990}. A representative set of literature which works on finding these solutions for various flow can be found in Refs.~\cite{waleffe2001,waleffe2009,kawahara2001,viswanath2007,chandler2013,willis2017}. These nonlinear solutions are often referred to as Exact Coherent Structures (ECSs) or Recurrent Flows and are supposed to play the same role of UPOs of low-dimensional systems in shaping the structure of turbulent motion. The exact nature and significance of ECSs has not yet been fully understood \cite{kawahara2012, graham2021}, but there is evidence that they are repeatedly shadowed by turbulent trajectories \citep{suri2020,krygier2021,crowley2022}.
Attempts to apply cycle averaging formulae to predict the statistical properties of turbulence from ECSs have also been made \citep{chandler2013, page2024}. These have, however, faced the challenge that identifying all the structures required for the cycle expansion is not straightforward and is prohibitively expensive using available numerical methods \cite{davidchack1999,diakonos1998,lan2004, viswanath2007, fazendeiro2010, schneider2022, schneider2023}. Further, as the dimension of the turbulent attractor increases with the Reynolds number, the likelihood that a turbulent trajectory shadows an ECS over its entire period decreases, thus impacting the quality of initial guesses generated using recurrence analysis techniques \citep{page2020}.

Given the difficulties in identifying a complete hierarchy of structures, we proposed in previous work \cite{lasagna2020} an alternative heuristic approach whereby computational resources are spent on identify one or few structures having a long time period \cite{saiki2009}. Such solutions may span a relatively large fraction of the attractor, visiting the neighbourhoods of a variety of relevant dynamical states. They may thus provide a good approximation of the statistical properties of turbulence, such as averages and probability distributions, potentially as accurate as that obtained from a rigorous but incomplete hierarchy of solutions and lifting the technical burden of having to determine the weights for the cycle expansion \citep{chandler2013,lucas2015}. Owing to the temporal periodicity of these simpler objects, adjoint methods for time-periodic systems \citep{hwang2008,giannetti2010,sierra2021} may be used, despite the instability, to obtain sensitivities of statistical properties with respect to problem parameters of an external forcing \citep{lasagna2018}. This information obtained from UPOs may then be leveraged to design flow control strategies \citep{meliga2016,giannetti2019} able to successfully manipulate the turbulent state.

For low-dimensional systems, this programme was carried out by first identifying UPOs with period hundreds or thousands times longer than the shortest UPO. These were found using a global Newton-Raphson search method \citep{lan2004,lasagna2018} that is insensitive to the marked sensitivity of such long chaotic trajectories to small perturbations that affects commonly-used shooting techniques. It was shown that period averages calculated on these UPOs appeared to converge to the long-time average of chaotic trajectories. Floquet exponents of such solutions, being the period averages of the local rate of growth of infinitesimal perturbations, also exhibited the same behaviour and appeared to converge to the Lyapunov exponents of the system calculated using standard methods \cite{benettin1980}. Nevertheless, extending this programme to high-dimensional fluid systems does not seem feasible at present because of the above-mentioned objective difficulties in locating ECSs, let alone solutions with long period, even using more recent search methods \citep{schneider2022,schneider2023}. If, however, the purpose is to obtain approximations of statistical properties and their sensitivity to problem parameters, the question is whether such long solutions really need to be determined exactly or whether it might be computationally advantageous to accept some violation of the governing equations as long as the resulting ``quasi-trajectories'' provide a sufficiently detailed ``sketch'' of the attractor whilst, of course, remaining physically relevant. The approach discussed in this paper to construct such quasi-trajectories parallels the ideas recently proposed in \citet{mccormack2024} and utilises dimensionality reduction as a strategy to trade-off the accuracy of statistical predictions with computational costs required to obtain them. Dimensionality reduction is motivated by the well-established notion that fluid systems display low-dimensional characteristics, and that the attractor lives in a low-dimensional subspace \citep{foias1988}. The dimensionality reduction used in this work demonstrates that the Lorenz attractor also lives in a low-dimensional subspace of the total state-space. Utilising the dimensionality reduction here, although not as pronounced as what can found for fluid systems, provides the framework that can be applied to the Navier-Stokes equations.

In this work, resolvent analysis in the formalism described in Ref.~\cite{mckeon2010} is utilised. Resolvent analysis is an operator-based model reduction technique that has been utilised extensively for the purpose of analysis, control, and modelling of fluid flows \cite{gomez2016,beneddine2017,jin2021,gayme2019,garnaud2013,gayme2010,symon2018}, and has been shown to provide efficient low-dimensional representations of ECSs found for pressure-driven pipe flow and plane Couette flow \cite{sharma2016}, with the theoretical correspondence between resolvent modes and invariant solutions to the Navier-Stokes equations being shown in Ref.~\cite{mezic2016}. Resolvent analysis is used here to provide a hierarchy of temporal basis functions onto which the dynamics can be projected, with the assumption that these are able to capture the majority of the dynamics of the turbulent flow with fewer dynamical variables. These basis functions are especially suited to model time-periodic exact solutions, as a hierarchy of modes is obtained at each temporal frequency (and spatial wavenumber for problems with statistically homogeneous spatial directions), so that temporal periodicity is built in explicitly in the modal expansion. Quasi-trajectories are then generated by adapting the variational optimisation methodology developed in Refs.~\cite{fazendeiro2010,farazmand2016,schneider2022} to the present low-dimensional settings. The method is equivalent to optimising an objective that measures the violation of the projected governing equation by a candidate quasi-trajectory. The dimensionality reduction restricts the problem to only the space of solutions that is most dynamically significant, eliminating superfluous search directions of the optimisation and leading to computational savings. Previous attempts at utilising resolvent analysis to construct approximate solutions of the governing equations were made in Ref.~\cite{barthel2021}, although steady solutions of Taylor-Couette flow were sought for, while here we consider time-dependent solutions that model dynamically-relevant processes. In addition, resolvent analysis has been used for a "project-and-search" algorithm \cite{sharma2020} that sought new ECSs by projecting known solutions onto a reduced set of resolvent modes. These projected solutions are then used as the initial condition for a new search, which was then successful in finding new equilibria and periodic orbits that had not yet been observed in the literature. 

The main contribution of this paper is to lay out the mathematical details of the framework used to generate time-periodic quasi-trajectories with the resolvent-based modelling, showing how variational optimisation methods and resolvent analysis can be combined. For the sake of providing a proof of concept on a prototypical chaotic system, the methodology is applied to the Lorenz 1963 system \cite{lorenz1963}. Despite being a low-dimensional problem, this system is chosen as it is a computationally accessible test-bed to make a first assessment of the numerical properties of the proposed framework. The application to fluid flows, and a detailed analysis of the role of the modal selection and truncation on the characteristics of the quasi-trajectories obtained is left to future work. The framework is introduced in \sref{sec:methodology}. In \sref{sec:lorenz} the methodology is applied to the Lorenz 1963 system \cite{lorenz1963}, with results reported in \sref{sec:lorenz-results} focusing on how statistics of observables obtained from quasi-trajectories compare to those obtained from long chaotic simulations. To conclude, the findings of the paper are summarised, and the future work is discussed in \sref{sec:conclusions}.

\section{Methodology}\label{sec:methodology}
In this section, the main components of the methodology are described. First the variational optimisation is introduced, showing how it can construct exact periodic solutions. This is followed with an explanation of the resolvent analysis, showing how resolvent modes are generated, what they represent, and how they can be truncated to allow the construction of a reduced-order model. This leads to an explanation of how the resolvent modes can be used as a basis for a low-dimensional projection of the dynamics, restricting the variational optimisation to the resolvent subspace. The section is then concluded with the numerical details of the implementation, explaining how the resolvent-based optimisation is constructed and how the main dynamical quantities are computed.

\subsection{Variational Optimiser}\label{sec:variational-solver}
To construct quasi-trajectories, an optimisation approach is used. In essence, this methodology aims to minimise an objective that measures the violation of the governing equation by a given state space trajectory. Although the framing is different, this approach is exactly equivalent to the adjoint solver methodology described in Ref.~\cite{farazmand2016}.

Consider the general autonomous dynamical system defined with the evolution equation
\begin{equation}
    \dv{\bm{x}}{t}=\bm{g}\left(\bm{x}\right),\quad\bm{x}\in\mathcal{M}\subseteq\mathbb{R}^d,
    \label{eq:gov-eq}
\end{equation}
\noindent where $\mathcal{M}$ is the state space (or phase space) for the system. The vector field $\bm{g}$ is assumed to be smooth, which implies that any solution to \eref{eq:gov-eq} is also smooth. The variational optimiser aims to find periodic solutions to \eref{eq:gov-eq}, such that the following condition is satisfied
\begin{equation}
    \int_t^{t+T}\bm{g}\left(\bm{x}\left(t^\prime\right)\right)\mathrm{d}t^\prime=0,\quad\forall t\in\mathbb{R}^{\geq0},\,T>0,
    \label{eq:periodic-constraint}
\end{equation}
\noindent where $T$ is the period of the solution. In general $T$ is not known a priori and should be included as part of the optimisation. We define the scaled time as $s=2\pi t/T=\omega t$, which is useful for decoupling the variation due to changes in shape of the trajectory and changes in its period. The variable $\omega=2\pi/T$ is called the fundamental frequency and corresponds to the smallest frequency that can be permitted in a finite time period. The trajectory $\bm{x}$ can now be expressed as a function over $s\in\left[0,\,2\pi\right)$. Additionally, the time derivative can now be expressed in terms of the scaled time as $d/dt=\omega d/ds$.

To help in the characterisation of the problem, the space of closed state space loops is defined as
\begin{equation}
    \mathscr{P}=\left\{\bm{x}\left(s\right)|\,\bm{x}\left(0\right)=\bm{x}\left(2\pi\right)\right\}.
\end{equation}

Thus, a trajectory is in $\mathscr{P}$ if and only if it is periodic. If a trajectory $\bm{x}$ and a given fundamental frequency $\omega$ satisfy \eref{eq:periodic-constraint}, then $\bm{x}\in\mathscr{P}$. Thus, there is a subset of $\mathscr{P}$ that represents exact periodic solutions to \eref{eq:gov-eq}. The general approach is to consider a particular initial loop $\bm{x}_0\in\mathscr{P}$ with a fundamental frequency $\omega_0$, and then modify both $\bm{x}_0$ and $\omega_0$ according to some update law such that $\bm{x}_n\in\mathscr{P}$ and the limit of $\bm{x}_n$ and $\omega_n$ is a solution to \eref{eq:gov-eq}

\begin{figure}
    \centering
    \includegraphics[width=0.36\textwidth]{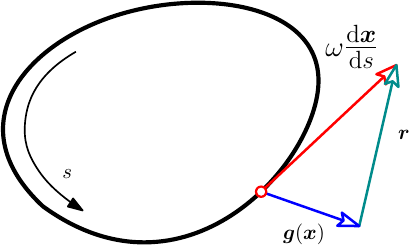}
    \caption{A state space loop that does not satisfy the governing equations, and thus has a non-zero residual $\bm{r}$.}
    \label{fig:local-residual}
\end{figure}

We define an inner-product on the space $\mathscr{P}$ as follows
\begin{equation}
    \innprod{\bm{x}}{\bm{y}}:=\int_0^{2\pi}\bm{x}\cdot\bm{y}\;\mathrm{d}s,
\end{equation}
\noindent which induces the norm $\norm{\bm{x}}=\sqrt{\innprod{\bm{x}}{\bm{x}}}$. Then we define a local residual
\begin{equation}
    \bm{r}:=\omega\dv{\bm{x}}{s}-\bm{g}\left(\bm{x}\right),
    \label{eq:lr}
\end{equation}
\noindent which is a measure of the local violation of the trajectory $\bm{x}$ with respect to the governing equations. \fref{fig:local-residual} is a schematic for what the local residual represents at each location in state space, given by the vector bridging the distance between the tangent vector to the state space loop ($\dv*{\bm{x}}{s}$) and the vector field $\bm{g}$. We also note that if $\bm{x}\in\mathscr{P}$ then it is true that $\bm{r}\in\mathscr{P}$. Thus, the problem can now be understood as finding some way to search for the trajectory $\bm{x}$ within the space $\mathscr{P}$ such that $\norm{\bm{r}}=0$, which is only true if $\bm{r}\left(t\right)=0$ for all $t\in\left[0,\,T\right)$. This motivates the definition of the global residual
\begin{equation}
    \mathcal{R}\left[\bm{x},\,\omega\right]:=\frac{1}{2}\norm{\bm{r}}^2,
    \label{eq:gr}
\end{equation}
\noindent as the measure of global violation of the governing equations by $\bm{x}$. By minimising $\mathcal{R}$ the alignment between the tangent vector $\dv*{\bm{x}}{s}$ and the governing vector field $\bm{g}$ is maximised. Therefore, if $\bm{x}\in\mathscr{P}$ is a solution to \eref{eq:gov-eq} for a given period $T$ then $\mathcal{R}=0$; otherwise $\mathcal{R}>0$.

The optimisation problem to find exact periodic solutions can now be summarised as follows
\begin{equation}
    \min_{\bm{x}\in\mathscr{P},\,\omega}\;\mathcal{R}\left[\bm{x},\,\omega\right].
    \label{eq:opt-prob}
\end{equation}

To solve this optimisation problem, it is desirable to have access to the gradient information on $\mathcal{R}$ with respect to variations in $\bm{x}$ and $\omega$. Variational calculus provides the tools to derive the functional derivative of $\mathcal{R}$ with respect to a given state space trajectory $\bm{x}$, giving the closed form expression
\begin{equation}
    \frac{\delta\mathcal{R}}{\delta\bm{x}}=-\omega\dv{\bm{r}}{s}-\bm{L}^\top\left(\bm{x}\right)\bm{r},
    \label{eq:res-grad}
\end{equation}
\noindent where $\bm{L}=\dv*{\bm{g}}{\bm{x}}$ is the Jacobian of $\bm{g}$ evaluated over the trajectory $\bm{x}$, and $\left(\cdot\right)^\top$ denotes the matrix transpose. This is a matrix ($\bm{L}\in\mathbb{R}^{d\times d}$) populated with the partial derivatives of the components of $\bm{g}$ with respect to the components of $\bm{x}$. The derivative with respect to $\omega$ is given as
\begin{equation}
    \pdv{\mathcal{R}}{\omega}=\innprod{\pdv{\bm{u}}{s}}{\bm{r}}.
    \label{eq:res-grad-freq}
\end{equation}

The details of the derivation of both \eref{eq:res-grad} and \eref{eq:res-grad-freq} are given in \aref{app:full-space-gradient}. Thus, by taking an initial state space loop and providing an optimisation algorithm with the objective $\mathcal{R}$ and gradients $\delta\mathcal{R}/\delta\bm{x}$ and $\pdv*{\mathcal{R}}{\omega}$, a monotonic decrease in $\mathcal{R}$ is guaranteed.

\subsection{Resolvent Analysis}\label{sec:resolvent-analysis}
The variational optimisation approach allows the construction of exact solutions to a set of dynamical equations. However, as discussed in \sref{sec:introduction} this is a costly and time-consuming procedure for high-dimensional systems encountered when dealing with spatio-temporally varying fields such as those produced by the Navier-Stokes equations. Resolvent analysis is introduced here as a way to construct low-order models of general dynamical systems that would allow for a projection onto a smaller subset reducing the dimensionality of the system while retaining important dynamical information \cite{sharma2016,sharma2013b,mckeon2013}. The formalism described in \cite{mckeon2010} has been specialised here for finite-dimensional systems.

Given a (not necessarily periodic) state space trajectory $\bm{x}$ with temporal length $T$, define the mean $\overline{\bm{x}}$ as follows
\begin{equation}
    \overline{\bm{x}}=\lim_{T\to\infty}\frac{1}{T}\int_0^T\bm{x}\left(t\right)\dd{t}.
    \label{eq:mean}
\end{equation}

Then the trajectory can be decomposed into the mean and fluctuations
\begin{equation}
    \bm{x}\left(t\right)=\overline{\bm{x}}+\bm{x}^\prime\left(t\right),
    \label{eq:decomposition}
\end{equation}
\noindent where $\bm{x}^\prime$ is the fluctuations of the trajectory. This decomposition can be substituted into \eref{eq:gov-eq}, noting that $\overline{\bm{x}}$ is invariant in time, to obtain an evolution equation for the state fluctuations
\begin{equation}
    \dv{\bm{x}^\prime}{t}=\bm{c}+\bm{L}\left(\overline{\bm{x}}\right)\bm{x}^\prime+\bm{f}\left(\bm{x}^\prime\right),
    \label{eq:gov-fluc}
\end{equation}
\noindent where $\bm{c}=\bm{g}\left(\overline{\bm{x}}\right)$, $\bm{L}\left(\overline{\bm{x}}\right)$ is the same Jacobian matrix as given in \eref{eq:res-grad} now evaluated only at $\overline{\bm{x}}$, and $\bm{f}$ is all the nonlinear terms in the expansion. Note that the constant $\bm{c}\neq0$ since the mean $\overline{\bm{x}}$ is not an equilibrium of the system.

Since we are concerned only with periodic solutions, it is natural to expand the trajectory as a Fourier series such that the condition $\bm{x}\in\mathscr{P}$ is automatically enforced, which is done as follows
\begin{equation}
    \bm{x}^\prime\left(s\right)=\sum_{n\in\mathbb{Z}^+}\left(\bm{x}^\prime_n+\text{c.c.}\right)e^{\iu ns}.
    \label{eq:fourier-exp}
\end{equation}
\noindent where the sum is over the positive integers $\mathbb{Z}^+=\left\{1,\,2,\,3,\dots\right\}$ and c.c. denotes the complex conjugate of each term. Since $\bm{x}^\prime$ is real valued, the resulting series possesses a Hermitian symmetry i.e. $\bm{x}^\prime_{-n}=\bm{x}^{\prime*}_n$, where $\left(\cdot\right)^*$ denotes the complex conjugate, allowing the sum to be expressed only over the positive frequencies. In addition, the zero frequency is not included in \eref{eq:fourier-exp} since the fluctuations by definition have zero mean component, that is, $\bm{x}_0=0$. The Fourier coefficients are related to the fluctuation with the identity
\begin{equation}
    \bm{x}^\prime_n=\frac{1}{2\pi}\int_0^{2\pi}\bm{x}^\prime\left(s\right)e^{-\iu ns}\;\mathrm{d}s.
    \label{eq:inverse-fourier}
\end{equation}

Expanding \eref{eq:gov-fluc} in terms of the Fourier components provides an algebraic equation governing the Fourier coefficients of the state fluctuations
\begin{subequations}
    \begin{align}
        \iu n\omega\bm{x}^\prime_n&=\bm{L}\left(\overline{\bm{x}}\right)\bm{x}^\prime_n+\bm{f}_n, \quad n\in\mathbb{Z}\setminus\left\{0\right\},\label{eq:gov-four} \\
        0&=\bm{c}+\bm{f}_0,\label{eq:gov-mean}
    \end{align}
    \label{eq:gov-eq-spectral}
\end{subequations}
\noindent where $\bm{f}_n$ are the Fourier coefficients of the nonlinear terms given by
\begin{equation}
    \bm{f}_n=\frac{1}{2\pi}\int_0^{2\pi}\bm{f}\left(\bm{x}^\prime\left(s\right)\right)e^{-ins}\;\mathrm{d}s.
\end{equation}

\eref{eq:gov-four} is the governing equation for the fluctuations' evolution. \eref{eq:gov-mean} is a constraint on the fluctuations imposed by the mean state, analogous to the Reynolds averaged Navier-Stokes equations, obtained through the same mean-fluctuation decomposition of a velocity field and averaging.

Rearranging \eref{eq:gov-four} to make the fluctuation coefficients the subject gives the following relationship with the nonlinearity
\begin{equation}
    \bm{x}^\prime_n=\bm{H}_n\bm{f}_n,\quad n\in\mathbb{Z}\setminus\left\{0\right\}.
    \label{eq:gov-res}
\end{equation}
\noindent where $\bm{H}_n\in\mathbb{C}^{d\times d}$ is the resolvent matrix, defined as
\begin{equation}
    \bm{H}_n=\left(\iu n\omega\bm{I}-\bm{L}\right)^{-1},\quad n\in\mathbb{Z}\setminus\left\{0\right\}
    \label{eq:resolvent-def}
\end{equation}

\eref{eq:gov-res} shows that the resolvent matrix acts as a transfer function relating the deviations of the system around the mean state due to the some forcing $\bm{f}_n$. This forcing term can be any general forcing in different contexts, however here it is known to represent the nonlinear interactions within the system that act to spread out the spectral content of the solution.

The next step in resolvent analysis is to perform a Singular Value Decomposition (SVD) on the resolvent matrix
\begin{equation}
    \bm{H}_n=\tilde{\bm{\Psi}}_n\tilde{\bm{\Sigma}}_n\tilde{\bm{\Phi}}_n^\dagger,
    \label{eq:svd}
\end{equation}
\noindent where $\tilde{\bm{\Psi}}_n\in\mathbb{C}^{d\times d}$ and $\tilde{\bm{\Phi}}_n\in\mathbb{C}^{d\times d}$ are the response and forcing resolvent matrices, respectively, and $\tilde{\bm{\Sigma}}_n\in\mathbb{R}^{d\times d}$ is the diagonal matrix of singular values, denoted with $\sigma_i$, ordered from largest to smallest. Here the $\left(\cdot\right)^\dagger$ represents the conjugate transpose of a matrix. Each column of $\tilde{\bm{\Psi}}_n$ and $\tilde{\bm{\Phi}}_n$ are a single response and forcing mode of the resolvent, which form a complete basis for the state of the system and the nonlinear forcing that drives it, respectively. The singular value associated with each mode pair provides the magnitude of response induced by the associated forcing, ordered by the size of said response. If the singular values in $\tilde{\bm{\Sigma}}_n$ decay quickly, i.e. $\sigma_i\gg\sigma_{i+1}$, then the system is most receptive to certain forcing modes in $\tilde{\bm{\Phi}}_n$, producing very large responses in the system in certain directions corresponding to the left-most columns of $\tilde{\bm{\Psi}}_n$. This means the resolvent, and the system as a whole, can be represented accurately with a truncated version of \eref{eq:svd} that removes the smaller singular values, as well as the forcing and response modes associated with them. Doing so it is possible to define the truncated SVD
\begin{equation}
    \bm{H}_n\approx\bm{\Psi}_n\bm{\Sigma}_n\bm{\Phi}_n^\dagger,
    \label{eq:svd_trunc}
\end{equation}
\noindent where $\bm{\Psi}_n\in\mathbb{C}^{d\times d_r}$, $\bm{\Phi}_n\in\mathbb{C}^{d\times d_r}$, and $d_r$ is the number of modes retained in each of the matrices. The right-most singular vectors of $\bm{\Psi}_n$ and $\bm{\Phi}_n$ have therefore been discarded along with their associated singular values. The degree of this truncation depends on the application. In the case of fluid dynamics and turbulence it has been shown that large separations of scale can occur between the singular values for certain wall-bounded flows \cite{mckeon2010}.

\subsection{Dimensionality Reduction}\label{sec:dimensionality-reduction}
As mentioned, the response matrix $\tilde{\bm{\Psi}}_n$ provide a complete basis for the state of the system. With the truncation established in \eref{eq:svd_trunc} it is now possible to use the reduced set of modes defined by the matrix $\bm{\Psi}_n$ as the basis for a projection that can reduce the dimensionality of the system, restricting the variational optimiser of \sref{sec:variational-solver} to the resolvent subspace. This projection onto the resolvent subspace is defined as
\begin{equation}
    \bm{a}_n=\bm{\Psi}_n^\dagger\bm{x}_n^\prime\quad\Leftrightarrow\quad\bm{x}_n^\prime=\bm{\Psi}_n\bm{a}_n.
    \label{eq:projection}
\end{equation}

\begin{figure}[t]
    \centering
    \includegraphics[width=0.2\textwidth]{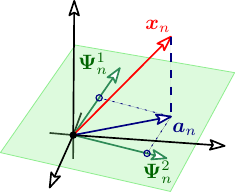}
    \caption{Schematic for the projection of the state onto the subspace defined by the truncated response modes.}
    \label{fig:projection_schematic}
\end{figure}

\fref{fig:projection_schematic} displays a schematic for this projection, showing that $\bm{a}_n\in\mathbb{C}^{d_r}$ in \eref{eq:projection} represents the best approximation to the state of the dynamical system within the subspace defined by the resolvent response modes.

To work in this new reduced space, it is necessary to modify the general optimisation problem given in \eref{eq:opt-prob} such that it can be expressed purely in terms of the resolvent subspace. A consequence of working in the reduced space defined by the modes in $\bm{\Psi}_n$ is that the mean state $\overline{\bm{x}}$ and the fundamental frequency $\omega$ are fixed, since they are a part of the derivation process of the modes themselves. They are required to be prescribed a priori and fixed over the duration of the optimisation of a quasi-trajectory. In principle, resolvent analysis can provide a basis for the mean, allowing for the optimisation to be performed over the whole spectrum. However, this is not implemented in practice since the mean is required regardless to construct the modes. Another consequence of working in the resolvent subspace, fixing the truncated modes $\bm{\Psi}_n$ and the mean $\overline{\bm{x}}$, is that it is possible that the residual may not have a zero. That is, it may not be possible to find an exact solution within the resolvent subspace defined by $\bm{\Psi}_n$ with the prescribed fundamental frequency $\omega$ and mean $\overline{\bm{x}}$. This means the solutions to UPOs is not possible in general given this methodology, however this is not a significant issue as the goal is construct periodic quasi-trajectories that only approximate the geometry of the attractor in state space. To find an exact UPOs, it is necessary to exactly know the mean of that particular UPO before starting the optimisation or allow the optimisation to make changes to the mean, provided some basis $\bm{\Psi}_0$, and the frequency. It is noted, however, that long UPOs have means that approach the mean obtained from chaotic solutions \cite{lasagna2020}, and therefore the chaotic mean can be used in the optimisation incurring only a small error. In regard to the fundamental frequency, it is known that as the period is increased the density of UPOs increases exponentially, thus for quasi-trajectories with very large periods it is known that solutions exist very close to the prescribed period and thus the convergence of the quasi-trajectory is not affected until small residuals are reached. In other words, if a small enough $\omega$ is chosen for a quasi-trajectory, i.e. the length of the trajectory is long, then the quasi-trajectory will be attracted close to a solution that exists with a very similar period.

In summary, the projection of the dynamics onto a the resolvent subspace defined by $\bm{\Psi}_n$ with a fixed $\overline{\bm{x}}$ and $\omega$ provides a reduced space on which the variational optimisation of \sref{sec:variational-solver} can be performed. In general, this subspace does not contain zeros of $\mathcal{R}$, representing UPOs of the system. Instead, the low-order model constructed, for large enough periods and using the approximate chaotic mean, can provide a good enough approximation of the dynamics to allow the construction of these quasi-trajectories that can output accurate statistics without being required to have $\mathcal{R}\approx0$.

Computing the global residual given a set of modal coefficients $\bm{a}_n$ can be done as follows. First, the definition of the global residual given in \eref{eq:gr} can be expressed in spectral space by substituting in the Fourier series of the local residual
\begin{equation}
    \mathcal{R}=\frac{1}{2}\bm{r}^\dagger_0\bm{r}_0+\sum_{n\in\mathbb{Z}^+}\bm{r}^\dagger_n\bm{r}_n,
    \label{eq:gr-spectral}
\end{equation}
\noindent where
\begin{equation}
    \bm{r}_n=\frac{1}{2\pi}\int_0^{2\pi}\bm{r}\left(s\right)e^{\iu ns}\;\mathrm{d}s.
    \label{eq:local-residual-spectral}
\end{equation}

The coefficients $\bm{r}_n$ are related to the state-vector as follows
\begin{subequations}
    \begin{align}
        \bm{r}_n&=in\omega\bm{x}^\prime_n-\bm{L}\left(\overline{\bm{x}}\right)\bm{x}^\prime_n+\bm{f}_n,\,n\in\mathbb{Z}\setminus\left\{0\right\}, \\
        \bm{r}_0&=\bm{c}+\bm{f}_0,
    \end{align}
    \label{eq:lr-spectral-def}
\end{subequations}
\noindent which are the Fourier transform of the governing equations as given in \eref{eq:gov-eq-spectral}. We remark that the equations for both the fluctuations and the mean are included in \eref{eq:gr-spectral}.

Left multiplying \eref{eq:lr-spectral-def} by $\bm{\Psi}^\dagger_n$ and using \eref{eq:projection} a reduced space local residual expression can be derived
\begin{subequations}
    \begin{align}
        \bm{\rho}_n&=in\omega\bm{a}_n-\bm{L}\left(\overline{\bm{x}}\right)\bm{a}_n+\bm{\Psi}^\dagger_n\bm{f}_n,\,n\in\mathbb{Z}\setminus\left\{0\right\}, \\
        \bm{\rho}_0&=\bm{\Psi}^\dagger_0\left(\bm{c}+\bm{f}_0\right),
    \end{align}
\end{subequations}
\noindent where $\bm{\rho}_n=\bm{\Psi}^\dagger_n\bm{r}_n$. Finally the global residual can then be computed with
\begin{equation}
    \mathcal{R}=\frac{1}{2}\bm{\rho}^\dagger_0\bm{\rho}_0+\sum_{n\in\mathbb{Z}^+}\bm{\rho}^\dagger_n\bm{\rho}_n,
\end{equation}
\noindent utilising the fact that $\bm{\Psi}^\dagger_n\bm{\Psi}_n=\bm{I}$. With this in mind, the resolvent-based optimisation problem can be stated as such
\begin{equation}
    \min_{\bm{a}_n,\,\forall n\in\mathbb{Z}\setminus\left\{0\right\}}\quad\mathcal{R}\left(\left\{\bm{a}_n\right\}\right).
    \label{eq:proj-opt-prob}
\end{equation}

In effect, the only change between \eref{eq:proj-opt-prob} and \eref{eq:opt-prob} is that the optimisation variables have been converted from the full-space trajectory $\bm{x}$ and $\omega$, to just $\bm{a}_n$. It should also be noted that the coefficient $\bm{a}_0$ is not included in the optimisation, since $\bm{a}_n$ is defined in \eref{eq:projection} in terms of the fluctuations $\bm{x}^\prime_n$. This means $\bm{a}_0=0$ by definition and is not included as part of the optimisation problem, which is a consequence of fixing $\overline{\bm{x}}$. The computational details of computing $\mathcal{R}$ are given in \sref{sec:numerical-details}.

In order to perform gradient-based optimisation within this resolvent subspace, an expression for $\partial\mathcal{R}/\partial\bm{a}_n$ is required. The gradient of $\mathcal{R}$ with respect to $\omega$ is not required in the reduced space as $\omega$ is fixed over the duration of the optimisation. Obtaining an expression for $\pdv*{\mathcal{R}}{\bm{a}_n}$ is done in two transformation steps. First Fourier transform \eref{eq:res-grad} to get the following expression
\begin{align}
    \frac{\partial\mathcal{R}}{\partial\bm{x}_n}&=\frac{1}{2\pi}\int_0^{2\pi}\frac{\delta\mathcal{R}}{\delta\bm{x}}e^{-\iu ns}\mathrm{d}s, \\
    &=-\iu n\omega\bm{r}_n-\left(\bm{L}\left(\bm{x}\right)^\top\bm{r}\right)_n.
    \label{eq:spectral-gradient}
\end{align}

Second, we project $\partial\mathcal{R}/\partial\bm{x}_n$ onto the resolvent subspace in the same way the state is projected to obtain
\begin{equation}
    \frac{\partial\mathcal{R}}{\partial\bm{a}_n}=\bm{\Psi}_n^\dagger\frac{\partial\mathcal{R}}{\partial\bm{x}_n},\quad n\in\mathbb{Z}\setminus0.
    \label{eq:gradient-projection}
\end{equation}

The proof of \eref{eq:spectral-gradient} and \eref{eq:gradient-projection} is given in \aref{app:gradient-projection}. Note that the mean frequency is not included due to the modes $\bm{\Psi}_n$ not being properly defined for the mean. This automatically ensures that using \eref{eq:gradient-projection} for a gradient-based optimisation does not modify the mean coefficient $\bm{a}_0$. As long as the optimisation is initialised with $\bm{a}_0=0$, the mean $\overline{\bm{x}}$ is guaranteed to be fixed.

\subsection{Numerical Details}\label{sec:numerical-details}

\begin{figure*}[t]
    \centering
    \includegraphics{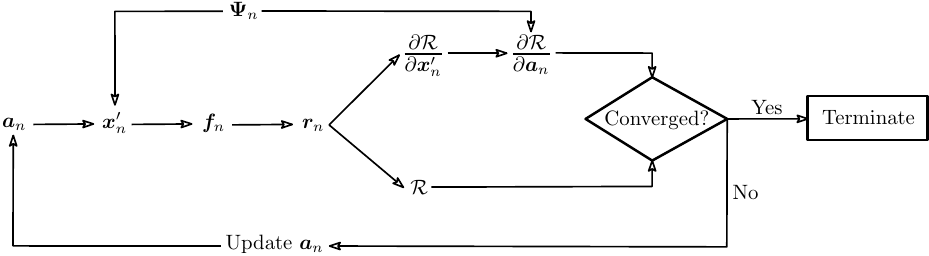}
    \caption{Flow diagram of a single iteration of the optimisation loop used to compute the residual and its gradient starting from the modal coefficients $\bm{a}_n$.}
    \label{fig:optimisation-loop}
\end{figure*}

The optimisation process for each iteration is visualised with the flow diagram given in \fref{fig:optimisation-loop}. The optimisation is initialised with a mean state $\overline{\bm{x}}$ and fundamental frequency $\omega$ that are fixed throughout the duration of the optimisation. These are used as inputs for resolvent analysis to generate the set of modes $\bm{\Psi}_n$ for $n\in\mathbb{Z}^+$. Next, the initial trajectory is generated. In the literature it is common to use close recurrences of chaotic trajectories obtained from direct simulations to initialise a given trajectory when trying to find exact periodic solutions \cite{cvitanovic1987, schneider2022}, which is particularly important for high-dimensional systems as the radius of convergence in such cases is small compared to the space in which the state can inhabit. In this work the results obtained were found to be very robust to the initial guess for the quasi-trajectory, and so the coefficients $\bm{a}_n$ were initialised randomly with a Gaussian distribution. Once $\overline{\bm{x}}$, $\bm{\Psi}_n$, and an initial $\bm{a}_n$ are known, the optimisation loop can begin.

When computing the residual and its gradient a ``pseudo-spectral'' method is used to reduce the time complexity of computing the nonlinear (typically quadratic) terms. This takes the form of expanding the coefficients $\bm{a}_n$ and inverse Fourier transforming the result to obtain the time domain representation of the quasi-trajectory. The nonlinear terms $\bm{f}\left(\bm{x}\right)$ can then be efficiently computed over the length of the quasi-trajectory, after which the result is Fourier transformed to obtain $\bm{f}_n$. Taking note of \fref{fig:optimisation-loop}, $\bm{f}_n$ is then used directly to compute $\bm{r}_n$. This means that over the course of the optimisation the trajectory is transformed from reduced to full space and then from spectral to time domain to be able to compute all the terms in the space where it is computationally most efficient. The result is the gradient $\partial\mathcal{R}/\partial\bm{x}^\prime_n$, which is then projected into the reduced space using \eref{eq:gradient-projection}, providing the required gradient with respect to the coefficients $\bm{a}_n$.

The choice of optimisation algorithm is an important aspect which has been relatively unexplored in the literature. As already mentioned, the choice of Refs.~\cite{farazmand2016,schneider2022,schneider2023} was to use an algorithm equivalent to gradient descent. This can be inefficient as the gradient descent has linear convergence as the minimum is approached \cite{nocedal2006}. For this reason, the L-BFGS optimisation algorithm is selected for this work \cite{liu1989,nocedal2011}. L-BFGS is a gradient-based quasi-Newton algorithm, first used to solve for periodic orbits in Ref.~\cite{otero2017}. L-BFGS included approximate Hessian information, significantly improves convergence rates, and generally requires fewer function evaluations compared to conjugate gradient methods, making a generally efficient method for gradient-based optimisation \cite{badreddine2014}. The L-BFGS iterations are coupled with a line search algorithm detailed in Refs.~\cite{nocedal2006,fletcher2000chapter2}.

The choice of convergence criteria is important in this context. Usually, a small global residual is used as it indicates that an exact solution to \eref{eq:gov-eq} has been found. In this work the focus is on quasi-trajectories for which the global residual is not expected to converge to a small value. Hence, it makes more sense here to track some relevant time-averaged observable of the system and terminate the iterations once this observable has converged to some value. This is explored more in \sref{sec:lorenz-results} using statistical measures of the Lorenz system.

\section{Application to the Lorenz System}\label{sec:lorenz}
The Lorenz system from Ref.~\cite{lorenz1963} will be used for the purpose of demonstrating the above methodology on a well known chaotic system. Even though it is an extensively studied problem, we demonstrate that the Lorenz equations permit an elegant, and to the best of the authors' knowledge not previously reported, dimensionality reduction when resolvent analysis is utilised, from three dimensions to two.

The governing equations are given as
\begin{subequations}
    \begin{align}
        \dv{x}{t}&=\sigma\left(y-x\right), \\
        \dv{y}{t}&=x\left(\rho-z\right)-y, \\
        \dv{z}{t}&=xy-\beta z.
    \end{align}
    \label{eq:lorenz-gov}
\end{subequations}

The standard parameter values of $\sigma=10$, $\rho=28$, and $\beta=\frac{8}{3}$ are used, for which it is known the system exhibits chaotic motion confined to a strange attractor. The governing equations are symmetric under the transformation $\left[x,\,y,\,z\right]\rightarrow\left[-x,\,-y,\,z\right]$, which implies that the mean has the form $\overline{\bm{x}}=\begin{pmatrix}0 & 0 & \overline{z}\end{pmatrix}^\top$ where $\overline{z}$ denotes the mean in the $z$-direction.

\subsection{Resolvent Derivation and Low-Oder Model}\label{sec:lorenz-resolvent}

By applying the mean-fluctuation decomposition of \eref{eq:decomposition} to the Lorenz system, the following evolution equation for the state fluctuations is obtained
\begin{subequations}
    \begin{align}
        \dv{x^\prime}{t}&=\sigma\left(y^\prime-x^\prime\right), \\
        \dv{y^\prime}{t}&=\left(\rho-\overline{z}\right)x^\prime-y^\prime-x^\prime z^\prime, \\
        \dv{z^\prime}{t}&=-\beta\left(\overline{z}+z^\prime\right)+x^\prime y^\prime.
    \end{align}
    \label{eq:lorenz-fluceq}
\end{subequations}

\eref{eq:lorenz-fluceq} can be expressed compactly as
\begin{equation}
    \dv{\bm{x}^\prime}{t}=\bm{c}+\bm{L}\left(\overline{\bm{x}}\right)\bm{x}^\prime+\bm{M}\bm{f}\left(\bm{x}^\prime\right),
    \label{eq:lorenz-compact}
\end{equation}
\noindent with $\bm{c}=\begin{pmatrix}0 & 0 & -\beta\overline{z}\end{pmatrix}^\top$ being the constant mean response of the system, the linearised Lorenz matrix evaluated at $\overline{\bm{x}}$ given as
\begin{equation*}
    \bm{L}\left(\overline{\bm{x}}\right)=\begin{pmatrix}
        -\sigma & \sigma & 0 \\
        \rho-\overline{z} & -1 & 0 \\
        0 & 0 & -\beta
    \end{pmatrix},
\end{equation*}
\noindent and the nonlinear influence matrix $\bm{M}$ defined as
\begin{equation}
    \bm{M}=\begin{pmatrix}
        0 & 0 \\
        -1 & 0 \\
        0 & 1
    \end{pmatrix}.
\end{equation}

The nonlinear forcing $\bm{f}\left(\bm{x}^\prime\right)=\begin{pmatrix}x^\prime z^\prime & x^\prime y^\prime\end{pmatrix}^\top$ is a two-element vector due to the fact that the first equation in \eref{eq:lorenz-gov} is linear.

By decomposing \eref{eq:lorenz-compact} into its Fourier modes and then rearranging, the resolvent operator for the Lorenz system is obtained
\begin{equation}
    \bm{H}_n=\left(\iu n\omega\bm{I}-\bm{L}\right)^{-1}\bm{M}=\begin{pmatrix}
        \alpha_n & 0 \\
        \beta_n & 0 \\
        0 & \gamma_n
    \end{pmatrix},
\end{equation}
\noindent where 
\begin{align}
    \alpha_n&=-\sigma/D_n, \\
    \beta_n&=-\left(\iu n\omega+\sigma\right)/D_n, \\
    \gamma_n&=1/\left(\iu n\omega+\beta\right), \\
    D_n&=\left(\iu n\omega+1\right)\left(\iu n\omega+\sigma\right)+\sigma\left(\overline{z}-\rho\right).
\end{align}

The matrix $\bm{M}$ reduces the size of the resolvent from a square matrix to a rank-$2$ rectangular matrix. The convenient structure allows an explicit expression for the SVD to be obtained as follows
\begin{align}
    \bm{H}_n&=\bm{\Psi}_n\bm{\Sigma}_n\bm{\Phi}_n^\dagger \\
    &=\begin{pmatrix}
        \zeta_n & 0 \\
        \eta_n & 0 \\
        0 & \kappa_n
    \end{pmatrix}\begin{pmatrix}
        \sigma_{1,n} & 0 \\
        0 & \sigma_{2,n}
    \end{pmatrix}\begin{pmatrix}
        1 & 0 \\
        0 & 1
    \end{pmatrix},
    \label{eq:lorenz-svd}
\end{align}
\noindent where the coefficients of the left singular matrix $\bm{\Psi}_n$ are given as $\zeta_n=\alpha_n/\sigma_{1,n}$, $\eta_n=\beta_n/\sigma_{1,n}$, and $\kappa_n=\gamma_n/\sigma_{2,n}$. The rank-$2$ nature of the resolvent means there are exactly $2$ singular values that govern the transfer of nonlinear forcing to the solution of the system. The response modes, defined by the columns of $\bm{\Psi}_n$, have the property that the first mode contains all the information from the $xy$-plane, while the second mode contains only the information in the $z$-direction. As such, retaining only a single pair of the modes restricts the dynamics to only the $xy$-plane or $z$-axis.

Due to the right singular vector, $\bm{\Phi}_n$, being equal to the identity matrix as shown in \eref{eq:lorenz-svd} the following expression for $\bm{\Sigma}_n$ can be derived
\begin{align*}
    \bm{\Psi}^\dagger\bm{\Psi}_n&=\bm{\Sigma_n^{-1}}\bm{H}_n^\dagger\bm{H}_n\bm{\Sigma}_n^{-1}=\bm{I}, \\
    \Rightarrow\bm{\Sigma}_n^2&=\bm{H}_n^\dagger\bm{H}_n,
\end{align*}
\noindent which gives for the individual singular values
\begin{align}
    \sigma_{1,n}&=\sqrt{\frac{\left(n\omega\right)^2+2\sigma^2}{|D_n|^2}},  \label{eq:singular-value1} \\
    \sigma_{2,n}&=\sqrt{\frac{1}{\left(n\omega\right)^2+\beta^2}}. \label{eq:singular-value2}
\end{align}

\begin{figure}[t]
    \centering
    \includegraphics{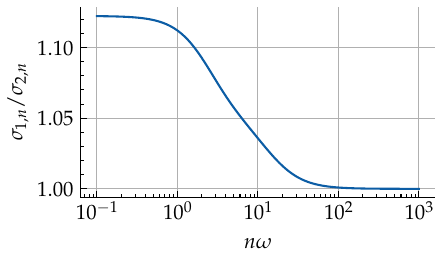}
    \caption{Ratio of the singular values for the Lorenz system plotted against the frequency.}
    \label{fig:singular-ratios}
\end{figure}

\fref{fig:singular-ratios} shows the ratio of the singular values as given in \eref{eq:singular-value1} and \eref{eq:singular-value2} as the frequency $n\omega$ is varied. There is no large separation of scale observed between these singular values and so the system cannot be accurately represented with a modal coefficients $\bm{a}_n\in\mathbb{C}^{d_r}$. Physically this is obvious if the response modes in \eref{eq:lorenz-svd} are inspected. If one of the response modes is neglected, then the dynamics is constrained to only the $xy$-plane or the $z$-axis (depending on which response modes is rejected), which cannot accurately reconstruct the structure of the strange attractor embedded in the full state space.

Therefore, the dimensionality reduction used in this work is from $\mathbb{C}^3$ to $\mathbb{C}^2$. This is an ``exact'' projection. That is to say, there has been no rejection of any non-zero singular values. This can be considered a special case of the more general (usually higher-dimension) setting, where there are more non-zero singular values as well as a distinctive separation of scales allowing the rejection of a finite number of relatively small singular values.

\subsection{Quasi-Trajectory Statistics}\label{sec:lorenz-results}

\begin{figure*}
    \centering
    \includegraphics{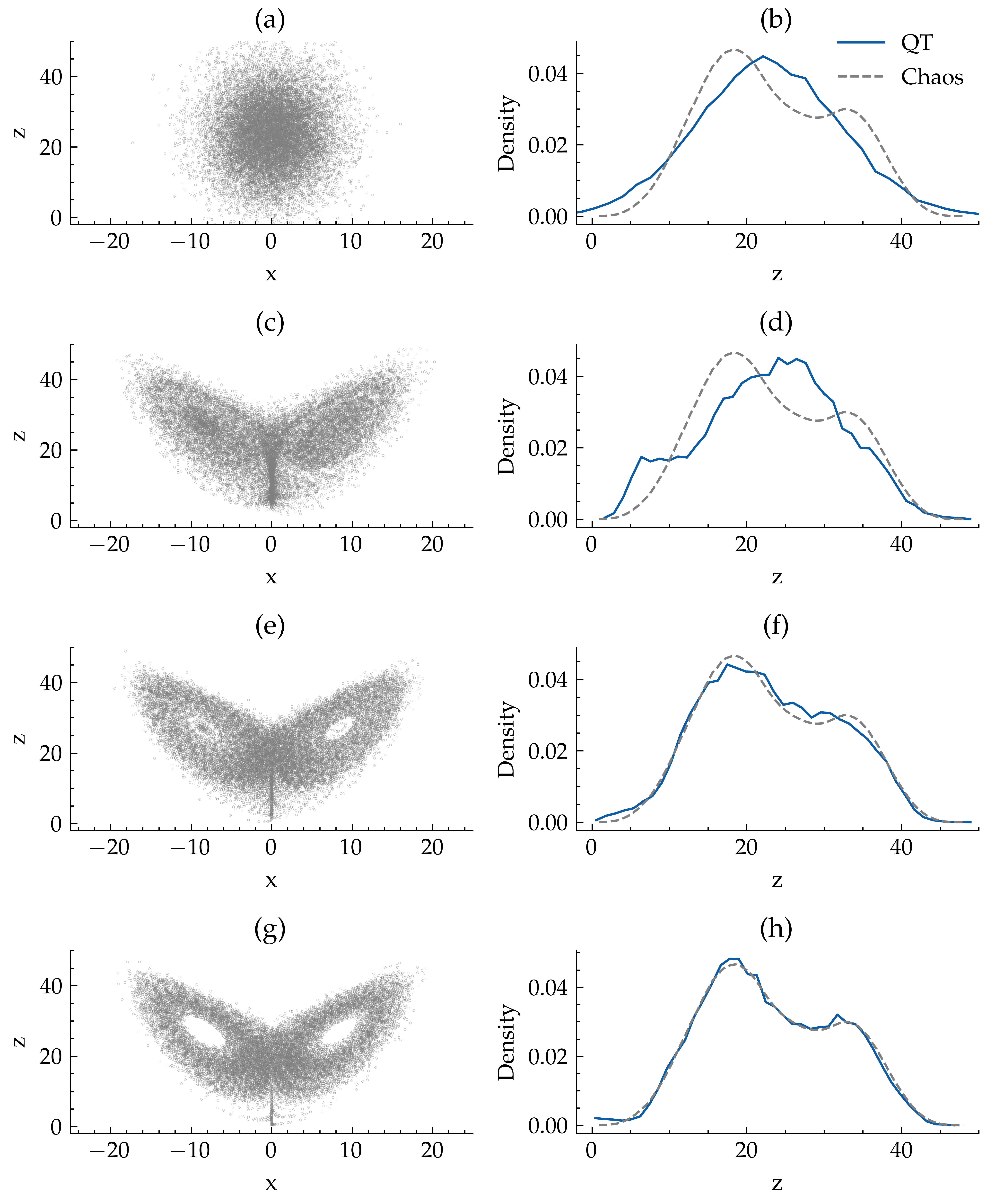}
    \caption{The optimisation of a quasi-trajectory, with the state space points shown in (a), (c), (e), and (g), and the probability distribution functions over the $z$-direction shown in (b), (d), (f), and (h), with the blue line corresponding to the Quasi-Trajectory (QT) and the dashed grey line corresponding to the distribution obtained from chaotic data. The initial trajectory (iteration $0$) is shown in (a), (b), iteration $100$ is shown in (c), (d), iteration $1000$ is shown in (e), (f), and iteration $10000$ is shown in (g), (h).}
    \label{fig:quasitrajectory-pdf}
\end{figure*}

In this section, the quasi-trajectories generated using the low-order resolvent-based model discussed in \sref{sec:lorenz-resolvent} for the Lorenz system are assessed in their ability to approximate the statistics of chaotic solutions. To facilitate this, the following observables are defined
\begin{align}
    \mathcal{J}_1\left(\bm{x}\right)&=\sqrt{x^2+y^2+z^2}, & \mathcal{J}_2\left(\bm{x}\right)&=xz.
    \label{eq:observables}
\end{align}
These observables will be averaged over the duration of a trajectory with period $T$, denoted by $\overline{\mathcal{J}_i}^T$.

\begin{figure*}
    \centering
    \includegraphics{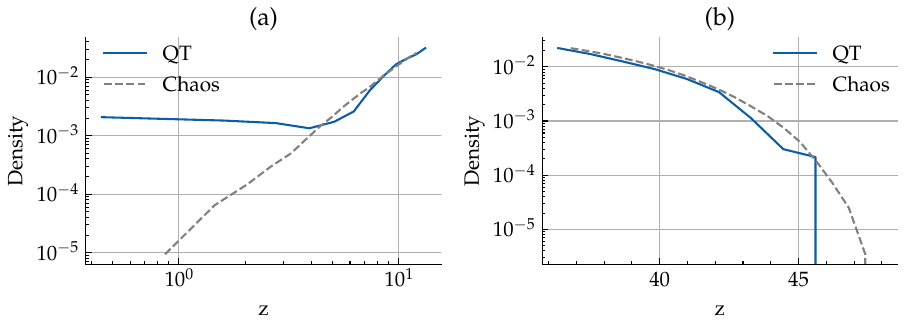}
    \caption{Tails of the PDF shown in panel (h) of \fref{fig:quasitrajectory-pdf}, panel (a) near $z=0$, and panel (b) near the extreme RHS of the distribution.}
    \label{fig:pdf-tails}
\end{figure*}

The data presented in this section is obtained from two sources. The first, denoted as ``chaos'' in the figures legends, is from chaotic simulations via black-box solvers supplied by the \emph{solve\_ivp} function from \emph{SciPy} with an explicit Runge-Kutta 45 method with adaptive time-stepping, described in \cite{dormand1980}, with the output trajectory being uniformly sampled in time. The second source is from quasi-trajectories. These are initialised randomly, generating the points in the time domain around the given mean $\overline{\bm{x}}$ with a standard deviation of $10$. The coefficients $\bm{a}_n$ are then determined from the time domain representation of the initial quasi-trajectory so that the optimisation can begin. The mean state used for the optimisation is set to $\overline{\bm{x}}=\begin{pmatrix}0 & 0 & 23.64\end{pmatrix}^\top$, obtained from independent chaotic simulations of the system using \eref{eq:mean}.

\fref{fig:quasitrajectory-pdf} displays the optimisation of a quasi-trajectory with a period of $T=1000$ starting from a random distribution around the mean and how it evolves over $100$, $1000$, and $10000$ iterations. On the left side the $xz$-projection of the quasi-trajectory sampled for $10000$ points is shown, reconstructed in the time domain from the coefficients $\bm{a}_n$. On the right side the corresponding probability distribution function (PDF) over the $z$-direction, with the solid line representing the PDF obtained from the quasi-trajectory and the dashed line being obtained from a chaotic solution. We first note that there is a qualitative resemblance between the quasi-trajectory and what is expected from a chaotic simulation of the Lorenz system. The noted resemblance is achieved after only roughly $1000$ iterations. This is a result of the optimisation seeking out the strange attractor very early, guiding the quasi-trajectory into a shape that lies on the attractor as well as possible for the given iteration. This is reinforced with the PDFs at each iteration, with the PDF shown in panels (f) and (h) of \fref{fig:quasitrajectory-pdf} agreeing well with the PDF obtained from a chaotic solution. Both the quasi-trajectory and chaotic solutions display a bimodal distribution. The PDF of the quasi-trajectory at each iteration is computed using $40$ bins over the range of $z$ values obtained by the quasi-trajectory. The coarseness of the bins used is due to the fact that as the number of bins is increased the PDF would display peaks that do not subside as the number of bins is increased. The presence of these peaks in the PDFs of periodic orbits was observed and discussed in Ref.~\cite{zoldi1998} and is a result of the turning points in the orbits.

There is a notable feature of the quasi-trajectory that is not present in chaotic trajectories. For the Lorenz system, there exists an unstable fixed point at the origin ($\bm{x}=\begin{pmatrix}0 & 0 & 0\end{pmatrix}$), which repels any trajectory that approaches it very quickly along its unstable manifold. However, as a consequence of the way in which the variational methodology is constructed, the residual is small around all fixed points regardless of their stability. This means that a quasi-trajectory is not heavily penalised for drifting away from the strange attractor towards the unstable fixed point at the origin. This can be observed in panels (g,h) of \fref{fig:quasitrajectory-pdf}, where there is a small increase in the density of the quasi-trajectory near the origin compared to the chaotic PDF that approaches zero as $z$ goes to zero. In panel (a) of \fref{fig:pdf-tails} a slice of the PDF of panel (h) from \fref{fig:quasitrajectory-pdf} is taken, plotted on a log-log scale, to show this trend more clearly. The chaotic PDF continues down as $z$ decreases in the fashion of a power law, while the quasi-trajectory PDF plateaus at a particular distance from $z=0$ after which it does not decrease any further. The effect on the statistics, however, is minimal. The quantitative effect of an unstable fixed point attracting quasi-trajectories under the variational optimisation is dependent on the stability characteristic of the particular fixed point. If the fixed point is close to neutrally stable, and therefore close to a bifurcation of some kind, it is difficult for the optimiser to converge towards it. The magnitude of attraction exerted on the optimiser is determined by the magnitude of the stability/instability of the point. There is also the possibility of so-called ``ghost'' points, discussed in Ref.~\cite{zheng2024}, which are fixed points that have bifurcated and no longer exist in the state-space, corresponding to a local minimum for the optimiser. Despite these fixed points no longer existing, there are remnants of their presence on the state-space which can have an effect on the optimisation, especially as the optimisers convergence is strongly affected by near neutral structures. In fluid problems, where there may be many more unstable fixed points scattered around the state-space, the exact effect this would have on the optimisation is not known and is a planned future topic of work.

Panel (b) of \fref{fig:pdf-tails} shows the right tail of the PDF from panel (h) of \fref{fig:quasitrajectory-pdf}, showing the probability of the quasi-trajectory to undergo a particularly large loop around a lobe of the strange attractor. It shows that the quasi-trajectory is able to capture some of the more unlikely/extreme events of the chaotic motion. There is an upper limit to the PDF that is smaller than that predicted from the chaotic PDF, however a larger quasi-trajectory would lead to extreme events at larger values of $z$ to be captured. With this in mind, extreme events can be important features in certain turbulent flows that display intermittency, and it is a topic for future work how well quasi-trajectories can capture them.

\begin{figure}[b]
    \centering
    \includegraphics{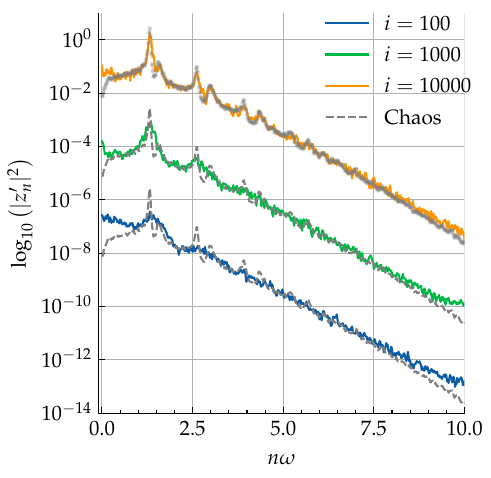}
    \caption{Comparisons of the power spectra obtained from the final quasi-trajectory in \fref{fig:quasitrajectory-pdf} at $100$, $1000$, and $10000$ iterations, and chaotic data. The spectra at each iteration (along with the corresponding chaotic spectra) is plotted offset from each other to improve readability.}
    \label{fig:power-spectra}
\end{figure}

\begin{figure*}
    \centering
    \includegraphics{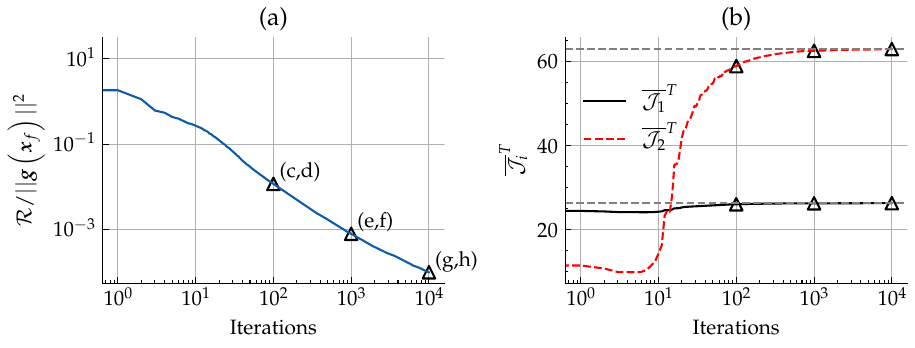}
    \caption{Trace plots of the global residual normalised by the size of the system response $\bm{g}$ quasi-trajectory at the end of the optimisation $\bm{x}_f$ (a) and the mean observables over the quasi-trajectory at each iteration of the optimisation (b), shown with the values obtained from a long chaotic trajectory as horizontal grey lines. Annotated on the plots are the iterations corresponding to the reconstructions shown in \fref{fig:quasitrajectory-pdf}.}
    \label{fig:trace-convergence}
\end{figure*}

In \fref{fig:power-spectra}, the power spectra of the final quasi-trajectory achieved in \fref{fig:quasitrajectory-pdf} at $100$, $1000$, and $10000$ iterations is compared with that obtained from the chaotic data. All spectra are obtained using Welch's method, with Hann windowing to reduce spectral leakage. Welch's method was used to compute the average power spectra of the quasi-trajectory. This is done since the raw spectrum obtained from each optimisation is not deterministic due to the random initialisation, which results in power spectra with seemingly random values. Using Welch's method averages this randomness out and reveals the deterministic statistics achieved by a given quasi-trajectory. This is fundamentally the same reason Welch's method is used when computing the power spectrum of chaotic signals \cite{andrews1988}. A similar trend is seen in \fref{fig:quasitrajectory-pdf}, in that the power spectra after $100$ does not particularly resemble the chaotic spectra, missing the important peak at $n\omega\approx1.3$ and its harmonics. After $1000$ iterations the quasi-trajectory spectra displays a spread out version of this peak with a couple of its harmonics, and finally $10000$ iterations shows the best agreement with the spectral peaks more clearly defined and multiple of its harmonics resolved in agreement with the chaotic data. The high frequency component of the spectra also gradually reduces over the duration of the optimisation, gradually approaching the spectral decay observed from the chaotic spectra. Notably, there is an increase in the spectral energy in the low frequencies that persists in the quasi-trajectory. This is possibly an artefact of the previously mentioned fixed point at the origin dragging part of the quasi-trajectory towards it. Near this point the quasi-trajectory moves rather slowly, approaching near a marginally unstable manifold, which adds an extra low frequency component to the spectra.

Panel (a) of \fref{fig:trace-convergence} the trace of the global residual, normalised by $\norm{\bm{g}\left(\bm{x}_f\right)}^2$ for the final quasi-trajectory obtained over the duration of the optimisation of the same quasi-trajectory as in \fref{fig:quasitrajectory-pdf}. It can be seen that the global residual has not yet converged to either a zero $\mathcal{R}=0$ or a non-zero minimum after $10000$ iterations. Nevertheless, the close qualitative resemblance observed in Figure \ref{fig:quasitrajectory-pdf} is achieved after only a moderate number of iterations. On the right of \fref{fig:trace-convergence} are the period-averaged observables defined in \eref{eq:observables} computed on the quasi-trajectory at each iteration, plotted with horizontal lines corresponding to the values of the mean observables computed from a long chaotic trajectory obtained with a numerical integration of the equations of motion. The values of the period averaged observables over the quasi-trajectory approach the chaotic values, displaying a convergence of the statistics well before the residual itself has converged, with most of the improvement being done between $10$ and $100$ iterations. Thus, it is reasonable to say that this quasi-trajectory has converged to the point of providing useful approximations to the statistics of the chaotic dynamics at around $1000$ iterations. 

The local residual $\bm{r}$ in equation \ref{eq:lr} can be viewed as a small perturbation imposed on the governing equations, and therefore a quasi-trajectory can be viewed as an exact solution to this slightly perturbed system. The ratio of the global residual to the norm of the system's right-hand side, $\mathcal{R}/\norm{\bm{g}\left(\bm{x}\right)}^2$, can then be viewed as a measure of the closeness of this forced system to original system. As such, these results show that nearby systems to the Lorenz system, or equivalently a lightly forced Lorenz system, have very similar statistics to each other. 

\begin{figure*}[t]
    \centering
    \includegraphics{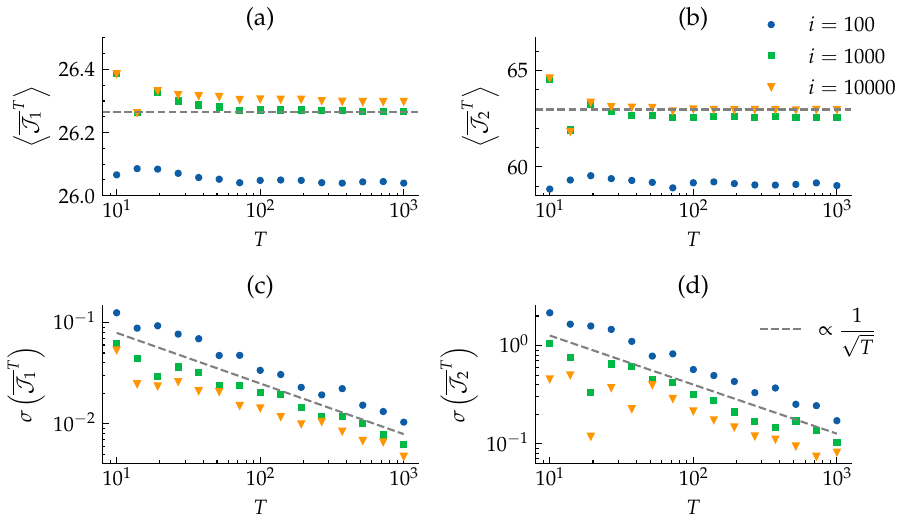}
    \caption{The ensemble average and standard deviations of the period averaged observables over a batch optimisation of $50$ quasi-trajectories, performed over a range of periods $T$. The optimisations were terminated at $100$, $1000$, and $10000$ iterations. The left side (panels (a,c)) show the observable $\overline{\mathcal{J}_1}^T$, and the right side (panels (b,d)) show the observable $\overline{\mathcal{J}_2}^T$. The top (panels (a,b)) show the ensemble averages, and the bottom (panels (c,d)) show the standard deviations.}
    \label{fig:observable-period}
\end{figure*}

\fref{fig:observable-period} shows the results of a number of batch optimisations at increasing periods $T$ performed for $100$, $1000$, and $10000$ iterations, with each batch consisting of $50$ quasi-trajectories. For reference, the shortest UPO of the Lorenz equations has a period of about 1.55 time units. Shown on the top are the ensemble averages of the period averaged observables within the batch, denoted by $\left\langle\overline{\mathcal{J}_i}^T\right\rangle$, and in the bottom is the associated standard deviation of the period averaged observables within each batch, denoted with $\sigma\left(\overline{\mathcal{J}_i}^T\right)$. The corresponding values for the period averaged observables obtained from a long chaotic trajectory are shown with the dashed grey lines. The trend for $1000$ and $10000$ is for the period averaged observables to approach the long chaotic values. The values of the period averaged observables for $100$ iterations exhibit poorer convergence towards the chaotic values, although the relative error is still rather small, being on the order of $1\%$ and $6\%$ for $\left\langle\overline{\mathcal{J}_1}^T\right\rangle$ and $\left\langle\overline{\mathcal{J}_2}^T\right\rangle$, respectively. The period averaged observables for the $10000$ iterations case are close to the long chaotic value even for the shortest periods shown. Panels (c,d) illustrate the change in the standard deviation of the period averaged observables with the quasi-trajectory period exhibiting a steady decline as the period increases. The rate of this decrease is roughly proportional to the inverse square root of the period shown in \fref{fig:observable-period} with the grey dashed line, which is a consequence of the central limit theorem. The larger period therefore produces quasi-trajectories that become more similar from a statistical point of view. The standard error of the estimation of $\sigma\left(\overline{J}^T_i\right)$ ranges from roughly $10\%$ to $0.1\%$ between the time-averaged observables \cite{lehmann2006}.
The result that the longer quasi-trajectories (for $T \gtrsim 20$) better reflect the statistics of the chaotic trajectories stem from their ability to explore the larger fractions of the strange attractor governing the chaotic dynamics. Therefore, a trade-off exists between the accuracy of the statistical predictions obtained and the speed at which the result can be achieved by varying the period of a quasi-trajectory. It should be noted that $\left\langle\overline{\mathcal{J}_1}^T\right\rangle$ approaches the chaotic value more closely for the optimisations that terminate at $1000$ iterations, whereas $\left\langle\overline{\mathcal{J}_2}^T\right\rangle$ is closer to the chaotic value for the optimisations that terminate at $10000$ iterations. This result suggests that certain observables may be most accurately captured during a quasi-trajectory optimisation.

\begin{figure*}[t]
    \centering
    \includegraphics{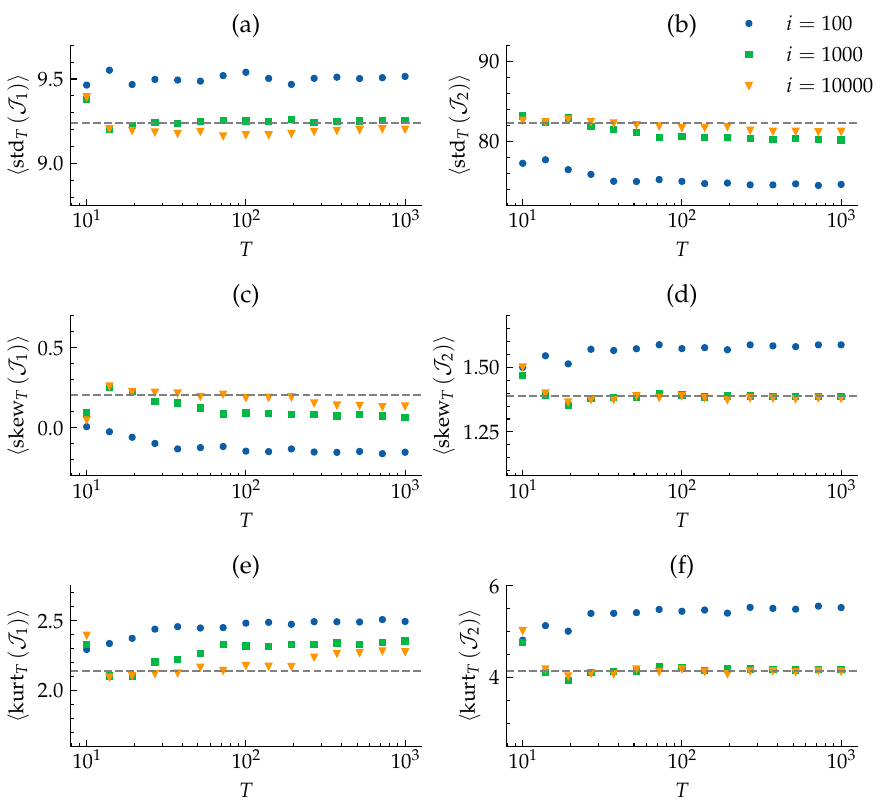}
    \caption{The ensemble average of the standard deviation (panels (a) and (b)), skewness (panels (c) and (d)), and kurtosis (panels (e) and (f)) of the observables $J_1$ and $J_2$ over the same set of batch optimisation as in \fref{fig:observable-period}.}
    \label{fig:observables-stats-with-period}
\end{figure*}

\fref{fig:observables-stats-with-period} is similar to \fref{fig:observable-period}, but instead shows the standard deviation (panels (a) and (b)), skewness (panels (c) and (d)), and kurtosis (panels (e) and (f)) of the observables taken over the period $T$ of a given quasi-trajectory, denoted with $\text{std}_T\left(\cdot\right)$, $\text{skew}_T\left(\cdot\right)$, and $\text{kurt}_T\left(\cdot\right)$ respectively. These statistical moments are then averaged over the ensemble of $50$ optimisations. The ensemble standard deviation follows the same descent trend as in \fref{fig:observable-period}, so is omitted for the sake of compactness. The expected trend is that for the same period quasi-trajectory, the higher-order statistical moments will be less accurate. \fref{fig:observables-stats-with-period} agrees with this, albeit with the estimation of the higher-order modes only degrading for the first observable, $J_1$. In fact, the quasi-trajectory statistics of $J_2$ appears to accurately reconstruct the statistical moments for periods $T\gtrsim 20$, implying the total distribution is faithfully reconstructed including the more extreme parts which would be emphasised by the kurtosis. The deviation of the moments of $J_1$ are difficult to directly attribute to a particular source, although it appears that more iterations leads to slightly better agreement with the value obtained from chaotic simulation. Additionally, the moments of $J_2$ appear to converge to a final value at a modest period, whereas the skewness and kurtosis of $J_1$ have seemingly not converged to the same degree. The optimisation then prioritises capturing certain aspects of the dynamics first, in this case balancing the nonlinear interactions over the total magnitude which takes larger periods and more iterations to accurately capture. In other words, the cross-correlations, represented by the observable $J_2$, are very accurately captured by quasi-trajectories with a modest period and without having converged the global residual.

\subsection{Time Cost of Computing Quasi-Trajectories}
In this small test case for the Lorenz system, the computation of the quasi-trajectories is much heavier than using a simple ODE solver. There are two primary reasons for this. The first is that the variational optimisation inherently scales worse than a time-stepping approach to solving the governing equations since the whole temporal evolution of the trajectory is considered simultaneously. The second, and more pertinent reason for this work, is the low-dimensionality of the system not admitting a large dimensionality reduction. For the algorithm implemented here, depicted in \fref{fig:optimisation-loop}, each iteration of the optimisation is dominated by the computation of the FFTs in full-space and the projection/expansion steps between full- and reduced-space. The time complexities of each of these operations are given by $\mathcal{O}\left(N\log\left(N\right)\right)$ and $\mathcal{O}\left(d_rN\right)$ respectively, where $N$ are the degrees of freedom of the system (original dimension of the system multiplied by the temporal modes used), and $d_r$ are the number of retained modes for the projection. For the variational optimiser to arrive at a sufficiently accurate estimate of the statistics more rapidly than a time-stepping method, it is required that the optimiser requires fewer degrees-of-freedom to accurately approximate the statistics, which can be best achieved through a large dimensionality reduction, i.e. $d_r\ll d$.

To illustrate for the Lorenz system, obtaining $1000$ optimisation iterations of a quasi-trajectory with a period of $T=20$ using $200$ temporal modes ($N=600$) with the dimensionality reduction $d_r=2$ takes $\mathcal{O}\left(1\right)$ seconds. Obtaining a chaotic ODE solution of the same period using \emph{solve\_ivp} takes $\mathcal{O}\left(0.1\right)$ seconds. In short, for low-dimensional systems, where the degrees of freedom are already small and the dimensionality reduction is not large, computing quasi-trajectories is more expensive than an ODE computation.

The potential power of the method becomes more apparent when applied to high-dimensional systems, specifically fluid turbulence, where the possible dimensionality reduction can be very large. This property, combined with the possibly smaller required degrees of the freedom in the full-space, and potentially relaxed time-step constraints, could reduce the time taken to find a statistically meaningful solution. The method described here performs the majority of the computations in the full state-space, only projecting back to the reduced space for the update to the optimisation state. It is possible to perform a complete optimisation loop within the reduced-space by converting the pseudo-spectral approach to computing the nonlinear terms to convolutions in the reduced-space, resulting in projected nonlinear terms which are computed in terms of the coefficients $\bm{a}_n$. Although convolutions are known to scale worse with the size of the computation, the dimensionality reduction could actually make the convolutions in reduced-space faster than FFTs in full-space. To finish this discussion, the global-in-time description of the flow that is implicit in the current framework would also permit temporal parallelism to be exploited, which may lead to a further speed-up that is not possible in direct time-stepping methods used by most flow solvers.

\section{Summary and Conclusions}\label{sec:conclusions}
In this work we proposed a modelling framework combining variational methods for the search of UPOs with resolvent analysis for dimensionality reduction. The fundamental idea is to construct candidate trajectories from a low-dimensional expansion of space-time basis functions with initially unknown amplitude coefficients. These coefficients are then found by optimising an objective function describing the overall violation of the governing equations along the trajectory. Using a reduced set of resolvent modes as a basis, where only the most receptive response modes (associated with the largest singular values of the resolvent matrix), the temporal periodicity of UPOs is built into the expansion. For the general case, the projection onto such a set of modes forces the resulting trajectories to live in a lower-dimensional space that rejects some of the dynamical information present in full state space. In addition, in using resolvent analysis the mean $\overline{\bm{x}}$ and period $T$ of the sought solution is fixed during the course of the optimisation. Thus, in general the governing equations can only be approximately solved. The Lorenz system, however, is a special case that allows for an exact projection of the $3$-dimensional dynamics onto a $2$-dimensional subspace that fully captures the dynamics. It was shown that no further truncation of the modes was possible while ensuring any accuracy of the resulting solutions, due to the two remaining singular values of the resolvent not having a large enough scale separation. Due to the truncation, solving this optimisation problem, may be computationally cheaper than identifying exact UPOs of the original system, especially for high-dimensional fluid systems. Despite this, these solutions, dubbed quasi-trajectories, may provide an adequate ``sketch'' of the attractor in state space and may thus have the ability to approximate to a sufficient degree of accuracy the statistical properties of the original system.

This framework is demonstrated on the Lorenz 1963 system, chosen as a computationally accessible test-bed for developing and testing the numerical techniques. For such system, resolvent analysis provides a natural dimensionality reduction that can be exploited to reduce the degrees of freedom of the system from three to two. Observables averaged over the period of a quasi-trajectory are monitored as the optimisation progresses. One key finding of this study is that such observables approach the values obtained from long chaotic trajectories rather quickly, after only $100$ - $1000$ iterations, with a great robustness to the initial condition used. This suggests that the governing equations projected on the low-dimensional subspace need not be solved exactly to obtain close approximations of the statistical properties of the original system. Instead, the gradient-based iteration procedure ``adjusts'' a candidate state space loop relatively quickly, pushing it towards the region of state space occupied by the attractor. The accuracy of the period averaged observables obtained from a quasi-trajectory increases with its period, with the variance in the value obtained from a given quasi-trajectory decreasing at the same time. This shows that longer quasi-trajectories better approximate the chaotic statistics, as a result of a larger portion of the attractor being explored.

The application of this framework to fluid systems remains to be explored. The dimensionality reduction afforded by resolvent analysis is much more pronounced in many flows of practical interest, compare to what can be obtained here for the Lorenz system. Here the dimensionality achieved is exact, losing no information in the projection step. This is not true for the fluid dynamics case since the mode truncation is generally motivated by a sufficiently large separation of scales between resolvent singular values. The expectation is that this truncation will not have a significant effect on the efficacy of the model, which is still required to be shown. The modal truncation could lead to large computational savings, relatively speaking, although its role on the quality of the approximation needs to be assessed, because there is no guarantee that the leading resolvent modes can capture the majority of the kinetic energy. This could result in approximations of the velocity field that miss important features. A second important task is to ascertain the impact of the attractor dimension on the convergence rate of observables computed from quasi-trajectories to the long-time statistics computed from chaotic trajectories. As shown, few iterations were sufficient here to obtain relatively good approximations of averages and probability distributions, but in a multi-scale problem such as turbulence the convergence may be slower. An answer to such question would then provide insight into how computational costs required to obtain estimates of the statistical properties depend on the problem dimension, as computational costs scale linearly with the number of iterations performed. These aspects are currently being considered and will be reported in future work.




\section*{Acknowledgments}
This paper is dedicated to the memory of Bruno Eckhardt. The work presented here was stimulated by discussions between B.E. and A.S. at the \emph{Recurrence, Self-Organization, and the Dynamics Of Turbulence} meeting, held at the Kavli Institute of Theoretical Physics, University of California, Santa Barbara in January 2017.

\appendix

\section{Residual Gradient}\label{sec:residual-gradient}

The gradient of the global residual $\mathcal{R}$ with respect to the state space loop $\bm{x}$ and its fundamental frequency $\omega$ is given in \eref{eq:res-grad} and \eref{eq:res-grad-freq}. Here these expressions are derived using variational calculus. In addition, the proof of the projection of the gradients onto the reduced subspace is provided.

\subsection{Full-Space Derivation}\label{app:full-space-gradient}

We will begin with \eref{eq:res-grad}. Let's, consider some perturbation to the state space loop $\bm{x}\rightarrow\bm{x}+\epsilon\bm{\delta x}$, such that $\left.\bm{\delta x}\right|_{t=0}=\left.\bm{\delta x}\right|_{t=T}$ and $\epsilon>0$ is some small real number. This perturbation propagates through all the variables that depend on the shape of the trajectory. The resulting perturbation in the local residual can be expressed as
\begin{equation}
    \begin{split}
        \bm{r}\left(\bm{x}+\epsilon\bm{\delta x}\right)&=\bm{r}+\bm{\delta r} \\
        &=\omega\dv{}{s}\left(\bm{x}+\epsilon\bm{\delta x}\right)-\bm{g}\left(\bm{x}+\epsilon\bm{\delta x}\right) \\
        &=\omega\dv{\bm{x}}{s}-\bm{g}\left(\bm{x}\right)+ \\
        &\qquad\epsilon\left(\omega\dv{}{s}\bm{\delta x}-\bm{L}\left(\bm{x}\right)\bm{\delta x}\right),
    \end{split}
\end{equation}
\noindent which when rearranged gives
\begin{equation}
    \bm{\delta r}=\epsilon\left(\omega\dv{}{s}\bm{\delta x}-\bm{L}\left(\bm{x}\right)\bm{\delta x}\right),
    \label{eq:lr-variation}
\end{equation}
\noindent where $\bm{L}$ is the Jacobian matrix of the vector-valued function $\bm{g}$ with respect to the state space variables $\bm{x}$. The perturbation of the global residual is given as
\begin{equation}
    \begin{split}
        \mathcal{R}\left[\bm{x}+\epsilon\bm{\delta x}\right]&=\frac{1}{2}\norm{\bm{r}+\bm{\delta r}}^2 \\
        &=\frac{1}{2}\norm{\bm{r}}^2+\innprod{\bm{r}}{\bm{\delta r}}+\frac{1}{2}\norm{\bm{\delta r}}^2 \\
        &=\frac{1}{2}\norm{\bm{r}}^2+ \\
        &\quad\epsilon\innprod{\bm{r}}{\omega\dv{}{s}\bm{\delta x}-\bm{L}\left(\bm{x}\right)\bm{\delta x}}+ \\
        &\quad\frac{\epsilon^2}{2}\norm{\omega\dv{}{s}\bm{\delta x}-\bm{L}\left(\bm{x}\right)\bm{\delta x}}^2.
    \end{split}
    \label{eq:gr-perturb}
\end{equation}

Now, variational calculus provides the following identities
\begin{equation}
    \left[\dv{\epsilon}\mathcal{R}\left[\bm{x}+\epsilon\delta\bm{x}\right]\right]_{\epsilon=0}=\innprod{\frac{\delta\mathcal{R}}{\bm{\delta x}}}{\bm{\delta x}},
    \label{eq:first-var}
\end{equation}

Applying \eref{eq:first-var} to the last equality of \eref{eq:gr-perturb} the following is obtained
\begin{equation}
    \innprod{\bm{r}}{\omega\dv{}{s}\bm{\delta x}-\bm{L}\left(\bm{x}\right)\bm{\delta x}}=\innprod{\frac{\delta\mathcal{R}}{\bm{\delta x}}}{\bm{\delta x}}.
    \label{eq:grad-relationship}
\end{equation}

Thus, to obtain a closed-form expression for $\fdv{\mathcal{R}}{\bm{x}}$ it is necessary to rearrange the left-hand side of \eref{eq:grad-relationship} such that it resembles the form of the right-hand side. Leveraging the bi-linearity of the inner-product, and performing integration by parts of the time derivative (noting that the boundary term disappears due to the periodicity of $\bm{\delta x}$) we get

\begin{equation}
    \innprod{-\omega\dv{\bm{r}}{s}-\bm{L}^\top\left(\bm{x}\right)\bm{r}}{\bm{\delta x}}=\innprod{\frac{\delta\mathcal{R}}{\bm{\delta x}}}{\bm{\delta x}},
\end{equation}
\noindent which, because $\bm{\delta x}$ is free to be any function we want, implies that
\begin{equation}
    \fdv{\mathcal{R}}{\bm{x}}=-\omega\dv{\bm{r}}{s}-\bm{L}^\top\left(\bm{x}\right)\bm{r}.
\end{equation}

Next, an expression for $\pdv*{\mathcal{R}}{\omega}$ can be obtained using standard calculus since $\omega$ is just a real number rather than a function like $\bm{x}$. Taking the definition of the global residual \eref{eq:gr}, substituting in \eref{eq:lr}, and then rearranging to make $\omega$ the subject, gives the following
\begin{equation}
    \begin{split}
        \mathcal{R}&=\frac{1}{2}\norm{\bm{r}}^2=\frac{1}{2}\norm{\omega\dv{\bm{x}}{s}-\bm{g}\left(\bm{x}\right)}^2 \\
        &=\frac{\omega^2}{2}\norm{\dv{\bm{x}}{s}}^2-\omega\innprod{\dv{\bm{x}}{s}}{\bm{g}\left(\bm{x}\right)}+\frac{1}{2}\norm{\bm{g}\left(\bm{x}\right)}^2.
    \end{split}
\end{equation}

Taking the derivative of this expression with respect to $\omega$ results in
\begin{equation}
    \pdv{\mathcal{R}}{\omega}=\omega\norm{\dv{\bm{x}}{s}}^2-\innprod{\dv{\bm{x}}{s}}{\bm{g}\left(\bm{x}\right)},
\end{equation}
\noindent which can be rearranged to a simpler form
\begin{equation}
    \pdv{\mathcal{R}}{\omega}=\innprod{\dv{\bm{x}}{s}}{\bm{r}}.
\end{equation}

\subsection{Projection onto the Resolvent Subspace}\label{app:gradient-projection}
The projection onto the reduced subspace is performed in two steps: first the expansion in terms of a Fourier series, and then a least-squares projection onto a subspace defined by the set of response modes obtained from resolvent analysis. First, to prove the expression given in \eref{eq:spectral-gradient}, we express the change in the global residual due to a change in the Fourier coefficients of $\bm{x}$ as
\begin{equation}
    \delta\mathcal{R}=\sum_{n\in\mathbb{Z}}\bm{\delta x}_n^\dagger\pdv{\mathcal{R}}{\bm{x}_n}.
\end{equation}

Thus, the gradient of $\mathcal{R}$ is expressed as
\begin{equation}
    \fdv{\mathcal{R}}{\bm{x}}=\sum_{n\in\mathbb{Z}}\fdv{\bm{x}_n}{\bm{x}}^\dagger\pdv{\mathcal{R}}{\bm{x}_n}.
    \label{eq:gradient-chain}
\end{equation}

To obtain an expression for the functional derivative of the Fourier coefficients $\bm{x}_n$ with respect to the trajectory $\bm{x}$, we use the definition \eref{eq:inverse-fourier} and introduce a perturbation $\epsilon\bm{\delta x}$
\begin{equation}
    \begin{split}
        \bm{x}_n\left(\bm{x}+\epsilon\bm{\delta x}\right)&=\frac{1}{2\pi}\int_0^{2\pi}\left(\bm{x}+\epsilon\bm{\delta x}\right)e^{-ins}\;\mathrm{d}s \\
        &=\bm{x}_n+\frac{\epsilon}{2\pi}\int_0^{2\pi}\bm{\delta x}e^{-ins}\;\mathrm{d}s.
    \end{split}
\end{equation}

Applying \eref{eq:first-var} to the above gives the following
\begin{equation}
    \fdv{\bm{x}_n}{\bm{x}}=\bm{I}e^{-ins}.
\end{equation}

Substituting this into \eref{eq:gradient-chain} gives
\begin{equation}
    \fdv{\mathcal{R}}{\bm{x}}=\sum_{n\in\mathbb{Z}}\pdv{\mathcal{R}}{\bm{x}_n}e^{\iu ns},
\end{equation}
\noindent which when the identity \eref{eq:inverse-fourier} is applied to provides
\begin{equation}
    \pdv{\mathcal{R}}{\bm{x}_n}=\frac{1}{2\pi}\int_0^{2\pi}\fdv{\mathcal{R}}{\bm{x}}e^{-ins}\;\mathrm{d}s=\left(\fdv{\mathcal{R}}{\bm{x}}\right)_n.
\end{equation}

The proof for \eref{eq:gradient-projection} is quite similar, if a little simpler. Using the chain rule, we have
\begin{equation}
    \pdv{\mathcal{R}}{\bm{a}_n}=\pdv{\bm{x}_n}{\bm{a}_n}^\dagger\pdv{\mathcal{R}}{\bm{x}_n}.
    \label{eq:gradient-projection-chain}
\end{equation}

Using the definition \eref{eq:projection}, the following derivative can be obtained
\begin{equation}
    \pdv{\bm{x}_n}{\bm{a}_n}=\bm{\Psi}_n,\quad n\in\mathbb{Z}\setminus\left\{0\right\}.
\end{equation}
\noindent which when substituted into \eref{eq:gradient-projection-chain} gives the final result
\begin{equation}
    \pdv{\mathcal{R}}{\bm{a}_n}=\bm{\Psi}_n^\dagger\pdv{\mathcal{R}}{\bm{x}_n},\quad n\in\mathbb{Z}\setminus\left\{0\right\}.
\end{equation}
\vspace{1pt}

\nocite{*}

\bibliography{LorenzPaper}

\end{document}